\documentclass[%
 reprint,
 amsmath,amssymb,
 aps,
prb,
]{revtex4-2}

\usepackage{xcolor}
\usepackage{graphicx}% Include figure files
\usepackage{dcolumn}% Align table columns on decimal point
\usepackage{bm}% bold math
\usepackage{afterpage}
\usepackage{hyperref}% add hypertext capabilities
\begin{document}

\preprint{APS/123-QED}

\title{Characterisation of temporal aiming for water waves with an anisotropic metabathymetry}% Force line breaks with \\

\author{Magdalini Koukouraki\thanks{Email:magdalini.koukouraki@espci.fr} }
 %\affiliation{Laboratoire de Physique et Mécanique des Milieux Hétérogènes, ESPCI Paris,
%PSL Research University, Sorbonne Université, Université Paris Cité,  75005 Paris, France}%Lines break automatically or can be forced with \\
%\altaffiliation[ magdalini.koukouraki@espci.fr]{\email{magdalini.koukouraki@espci.fr}}
\altaffiliation[Also at ]{Laboratoire de Physique et Mécanique des Milieux Hétérogènes, ESPCI Paris, PSL Research University, Sorbonne Université, Université Paris Cité,  75005 Paris, France}%Lines break automatically or can be forced with \\
\email{magdalini.koukouraki@espci.fr}
\author{Philippe Petitjeans}%
%\affiliation{%
 \affiliation{Laboratoire de Physique et Mécanique des Milieux Hétérogènes, ESPCI Paris,
PSL Research University, Sorbonne Université, Université Paris Cité,  75005 Paris, France}%Lines break automatically or can be forced with \\
% \altaffiliation[Also at ]{Laboratoire de Physique et Mécanique des Milieux Hétérogènes, ESPCI Paris,
%PSL Research University, Sorbonne Université, Université Paris Cité,  75005 Paris, France}%Lines break automatically or can be forced with \\
\email{phil@pmmh.espci.fr}
\author{Agnès Maurel}
\affiliation{Institut Langevin, ESPCI Paris, PSL Research University, 75005 Paris, France
}
\author{Vincent Pagneux}
\affiliation{%
Laboratoire d’Acoustique de l’Université du Mans, Le Mans Université, 72085 Le Mans, France
}%

\date{\today}% It is always \today, today,

\begin{abstract}
The deflection of waves by combining the effects of time modulation with anisotropy has been recently proposed in the context of electromagnetism. In this work, we characterise this phenomenon, called temporal aiming, for water waves using a time-varying metabathymetry. This metabathymetry is composed of thin vertical plates that are periodically arranged at the fluid bottom and which act as an effective anisotropic medium for the surface wave in the long-wavelength approximation. When this plate array is vertically lifted at the fluid bottom at a given time, the medium switches from isotropic to anisotropic, causing a wavepacket to scatter in time and deflect from its initial trajectory. Following a simple modelling, we obtain the scattering coefficients of the two waves generated due to the sudden medium change as well as the angle of deviation with respect to the incident angle. We then numerically evaluate this scattering problem with simulations of the full 2D effective anisotropic wave equation, with a time-dependent anisotropy tensor. Finally, we provide experimental evidence of the temporal aiming, using space time resolved measurement techniques, demonstrating the trajectory shift of a wavepacket and measuring its angle of deviation.

%\begin{description}
%\end{description}
\end{abstract}

\maketitle

\section{Introduction}
Time-varying metamaterials have in the past years provided unconventional ways to control and harness waves in various fields of wave physics, from photonics \cite{galiffi_photonics, tretyakov_tutorial, lurie_book, monticone, pena_opinion, engheta_science} to condensed matter \cite{stefanou_layered}, elastodynamics \cite{riva_elastic}, acoustics \cite{acoustics_TV} and water waves \cite{bacot,che2024generation}. The addition of time variation as an extra ingredient in the effective medium parameters, which was first examined in electromagnetism \cite{morgenthaler}, has opened up new possibilities for wave manipulation and unveiled interesting applications, such as time-reflection \cite{time_reflection_microwaves}, frequency conversion \cite{eleftheriadis_frequency_conversion}, parametric amplification \cite{pendry_amplification}, transient amplification  \cite{kiorpe_transient}, inverse prism \cite{caloz_inverse_prism} and antireflection temporal coatings \cite{pena_antireflection_coating}.

Temporal aiming is one interesting application in the realm of time-varying metamaterials and was first proposed in \cite{temporal_aiming} for electromagnetic waves as a means to guide a wavepacket in space by exploiting time variation. It was suggested to rapidly alter the permittivity tensor of the medium at a specific time $t=t_{0}$, inducing a switch in time from isotropic to anisotropic. In that way, the wavepacket encountering the new anisotropic medium at $t\geq t_{0}$ will be redirected and will travel along the direction of the energy flow which differs than the direction of the wave vector.  By making a transition back and forth between the two media at different time instants the direction of the wavepacket propagation is shifted multiple times, i.e. at each time that a medium switch occurs. Hence, the wavepacket is temporally guided in space without the use of spacial boundaries. The idea of temporal aiming has been supported with numerical simulations, but to the best of our knowledge there is no experimental demonstration yet. While this application is relevant for antenna communications and radars we believe that it can also be useful for other types of physical systems such as water waves.

In the field of water waves, wave control plays an important role for coastal and port protection, as well as various waterfront activities \cite{porter_cloaking_review, zhu2024controlling}. The bathymetric profile can significantly affect the wave dynamics and thus shaping it in different ways with the use of metamaterials can lead to unique wave phenomena. There has been a strong research activity in this spirit, dealing with scattering problems from a large variety of bathymetric shapes \cite{fitz_gerald, hamilton, evans_linton, porter_map, newman, porter_evans_vertical_barriers}, for instance structured bathymetries \cite{marangos_structured_bathymetry, maurel_pham}. In particular, spatially periodic microstructures such as periodic vertical plates have been long used as a means to engineer an effective medium \cite{papanicolaou, anisotropy}, whose parameters in the low frequency limit can be known by following homogenisation techniques \cite{porter_homogenisation}.  While the forementioned microstructures stand for space metamaterials, progress has been also made in implementing time-modulated media within the water wave framework, with a main challenge being in finding a mechanism that drives the time-variation without it acting as an additional source. For instance, time interfaces have been achieved using electrostriction, i.e. by applying an electric force on the water surface in order to modify the water wave speed \cite{apffel_frequency, apffel_disorder}. Moreover, vertically moving submerged plates of very small width also make excellent candidates to tackle wave phenomena linked to time variation, as they have been reported to be essentially source-free \cite{koukouraki_floquet, mei}. 

In this paper, we provide experimental evidence of the temporal aiming in the context of water waves by constructing the water wave analogue. This analogue holds in the classical linearised water wave theory \cite{mei, whitham, lamb, stoker} in the long wavelength approximation. The implementation of the temporal aiming for water waves involves using a metamaterial plate array placed at the fluid bottom with the ability to move vertically. More precisely, the plate array can be modelled by an effective 2D anisotropic medium in the long-wavelength limit, so that when it is abruptly lifted (or delifted) at a given time at the fluid bottom the medium switches from isotropic to anisotropic (or vice versa). First, we start by collecting the main idea behind the temporal aiming of water waves, combining the anisotropic metabathymetry with time variation. We briefly review wave propagation in the metabathymetry and discuss its dispersion relation characteristics. We then proceed with a basic modelling of the water wave scattering by a time interface, which yields the scattering coefficients for both an isotropic to anisotropic switch and vice versa, as well as the angle of deviation. The analytical relations are then compared with numerical results obtained by solving the full 2D anisotropic water wave equation in time using finite difference, where the effective bathymetry is modified everywhere in space at a specific time. Finally, we report on the experimental realisation of the temporal aiming and experimentally measure the angle of wavepacket deviation.
\section{Temporal aiming concept for water waves}
The temporal aiming requires two key ingredients: an anisotropic medium and a time interface. In this section we examine for water waves each one separately and discuss how to properly combine them for this purpose.
\subsection{Effective anisotropic medium in shallow water}\label{sA}
\begin{figure}[h!tbp]
\centering
\includegraphics[width=1 \columnwidth]{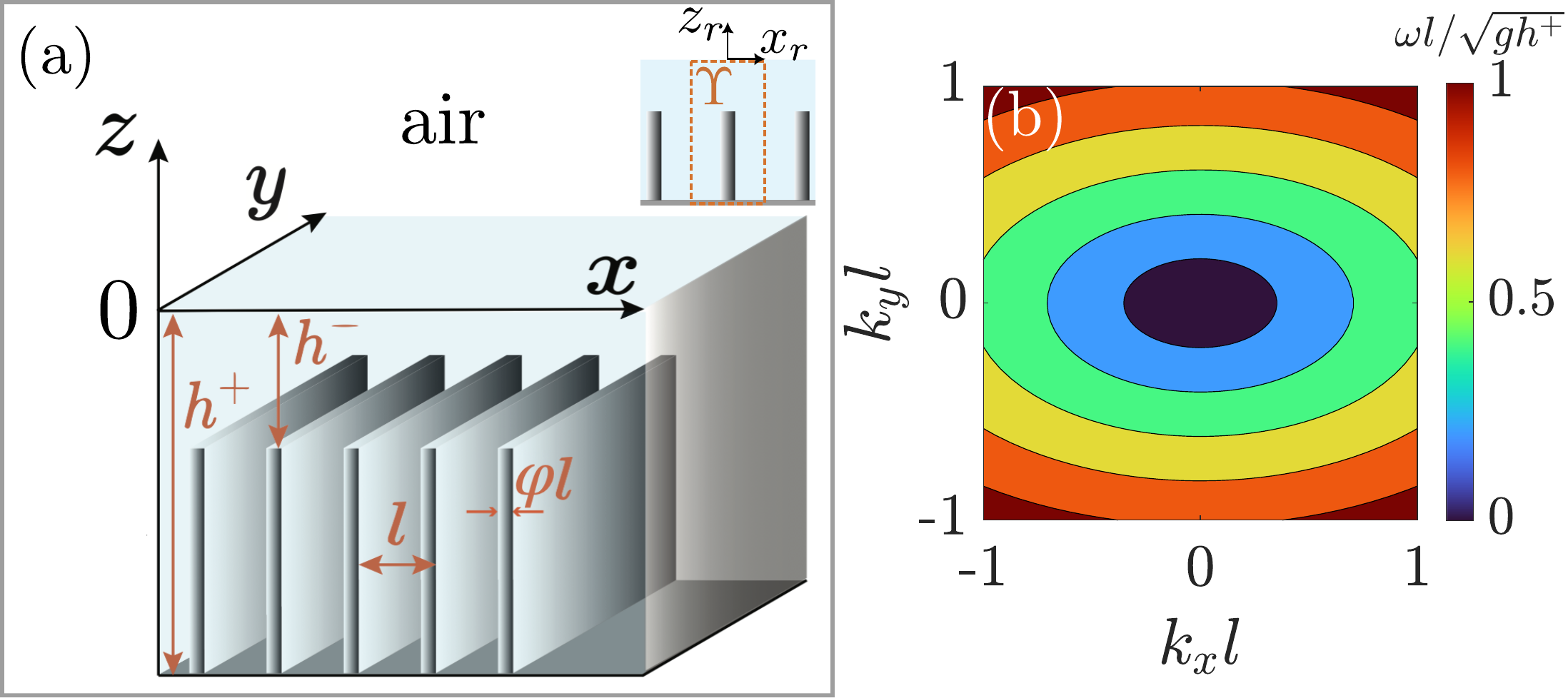}
\caption{(a) Plate array immersed in a water column, standing at the fluid bottom at position $z=-h^{+}$ and extending in height up to $z=-h^{-}$, while the water surface is located at $z=0$. Plates are periodic along $x$ with periodicity $l$ and width $\varphi l$, with $\varphi$ the filling fraction, and are uniform along $y$. Above the plate array a close-up view of the unit cell $\Upsilon$ is shown in the $(x_{r},z_{r})$ microscopic scale. (b) Anisotropic elliptic band structure of the effective medium for plate array characteristics $l=0.8~\mathrm{cm}$, $\varphi=0.0625$, $h^{+}=2~\mathrm{cm}$, $h^{-}=0.5~\mathrm{cm}$, and a range of dimensionless frequencies $\omega l/\sqrt{gh^{+}}\in [0,1]$.\label{effective_anis}}
\end{figure}

The classical water wave theory relies on the assumptions that the fluid is inviscid and incompressible and the flow is irrotational. In this paper, we consider this theory in the linear gravity wave regime, characterized by waves of small amplitude relative to both the water depth and wavelength, and where surface tension effects are negligible—i.e., for wavelengths larger than the capillary length. It has been shown \cite{papanicolaou, anisotropy} that a plate array sitting at the fluid bottom can create an effective anisotropic medium in the long-wavelength approximation, i.e. when the wavelength $\lambda$ is much larger than the array periodicity $l$. The equation describing wave propagation in this medium can be obtained by implementing a 3D homogenisation technique, yielding a 2D effective medium governed by
\begin{equation}\label{anis_eq}
\frac{\partial^{2}\eta}{\partial t^{2}}-g \nabla \cdot(\mathrm{H} \nabla \eta)=0, \quad \mathrm{H}=\left(\begin{array}{cc}
h_x & 0 \\
0 & h_y
\end{array}\right),
\end{equation}
with $\eta$ denoting the perturbed surface elevation and $g$ the gravitational acceleration. The effective water depths $h_x$ and $h_{y}$ are expressed as 
\begin{equation}\label{hx_relation}
\left\{
\begin{aligned}
& h_{x}=l\int_{\Upsilon} \frac{\partial \Phi}{\partial x_{r}}\mathrm{d}x_{r}\mathrm{d}z_{r},\\
& h_{y}=l[\varphi h^{-}+(1-\varphi) h^{+}],
\end{aligned} \right.
\end{equation}
with $\Phi$ the fluid velocity potential determined at the microscopic scale of the unit cell $\Upsilon$ with coordinates $(x_{r},z_{r})$, $h^{+}$, $h^{-}$ the water depths defined in Fig. \ref{effective_anis}(a) and $\varphi$ the filling fraction. Note that $h^{-}<h_{x}<h_{y}<h^{+}$. Hence, the 3D microstructure is replaced by the 2D effective anisotropic shallow water wave equation \eqref{anis_eq}. The dispersion relation of Eq. \eqref{anis_eq} reads as
\begin{equation} \label{elliptic_dispersion}
\omega^{2}= g\left(h_x k_{x}^{2}+h_y k_{y}^{2}\right),
\end{equation} 
with $\omega$ the radial frequency and where the wave vector has been written as $\boldsymbol{k}=k_{x} \hat{x}+k_{y} \hat{y}$, with $\hat{x}$ and $\hat{y}$ unitary vectors in the $x$ and $y$ axis. In the absence of plates, i.e. when $h^{+}=h^{-}$, the simple isotropic dispersion relation $\omega^{2}=gh^{+}k^{2}$ is recovered. 

When moving towards higher frequencies a dispersive correction can be included in the dispersion relation \eqref{elliptic_dispersion}, as discussed in \cite{anisotropy}. It consists of replacing the nondispersive Eq. \eqref{elliptic_dispersion} with the relation
\begin{equation}\label{dispersive_dispersion}
\omega^{2}= g\left(h_x k_{x}^{2}+h_y k_{y}^{2}\right)\tanh(kh)/kh,
\end{equation}
with $kh=\sqrt{k_{x}^{2}h_{x}^{2}+k_{y}^{2}h_{y}^{2}}$.

In Fig. \ref{effective_anis}(b), we illustrate the anisotropic elliptic band structure of the dispersion relation \eqref{elliptic_dispersion} with $l=0.8~\mathrm{cm}$, $\varphi=0.0625$, $h^{+}=2~\mathrm{cm}$, $h^{-}=0.5~\mathrm{cm}$. This elliptic dispersion relation tells us that there are two different effective water depths along $x$ and $y$. The anisotropy can be estimated from the ratio $\mathcal{A}\equiv \left( k_{x}/k_{y}\right)^{2}= h_{y}/h_{x}$, with a maximum anisotropy that is achieved for very thin plates that are tightly arranged, since in that case $h_{x}\rightarrow h^{+}$ and $h_{y}\rightarrow h^{-}$  (see \cite{anisotropy}). Furthermore, due to the anisotropic nature of the medium, the direction of the energy propagation (or the group velocity direction), forming an angle $\theta_{S}$ with the $x$-axis, differs from the wavevector orientation given by $k_{x}=k\cos\theta$ and $k_{y}=k\sin\theta$, where $\theta$ denotes the wavevector angle.  One can extract the analytical expression for $\theta_{S}$ by starting from the definition of the group velocity $\boldsymbol{c}_{g}=\partial \omega/\partial k_{x}\hat{x}+\partial \omega/\partial k_{y}\hat{y}$ and using Eq. \eqref{elliptic_dispersion} along with $\partial\omega/\partial k_{x}=c_{g}\cos\theta_{S}$, $\partial\omega/\partial k_{y}=c_{g}\sin\theta_{S}$. Hence, it follows that
 \begin{equation}\label{theta_S}
 \theta_S=\tan ^{-1}\left(\tan \theta \frac{h_y}{h_x}\right).
 \end{equation}
 One can clearly identify the analogy of Eq. \eqref{theta_S} with Eq. (2) of the paper \cite{temporal_aiming} for the angle of the Poynting vector in the case of anisotropic permittivity change. 
\subsection{Scattering by a time interface: Simple modelling}\label{sB}
\begin{figure}[h!tbp]
\centering
\includegraphics[width=1 \columnwidth]{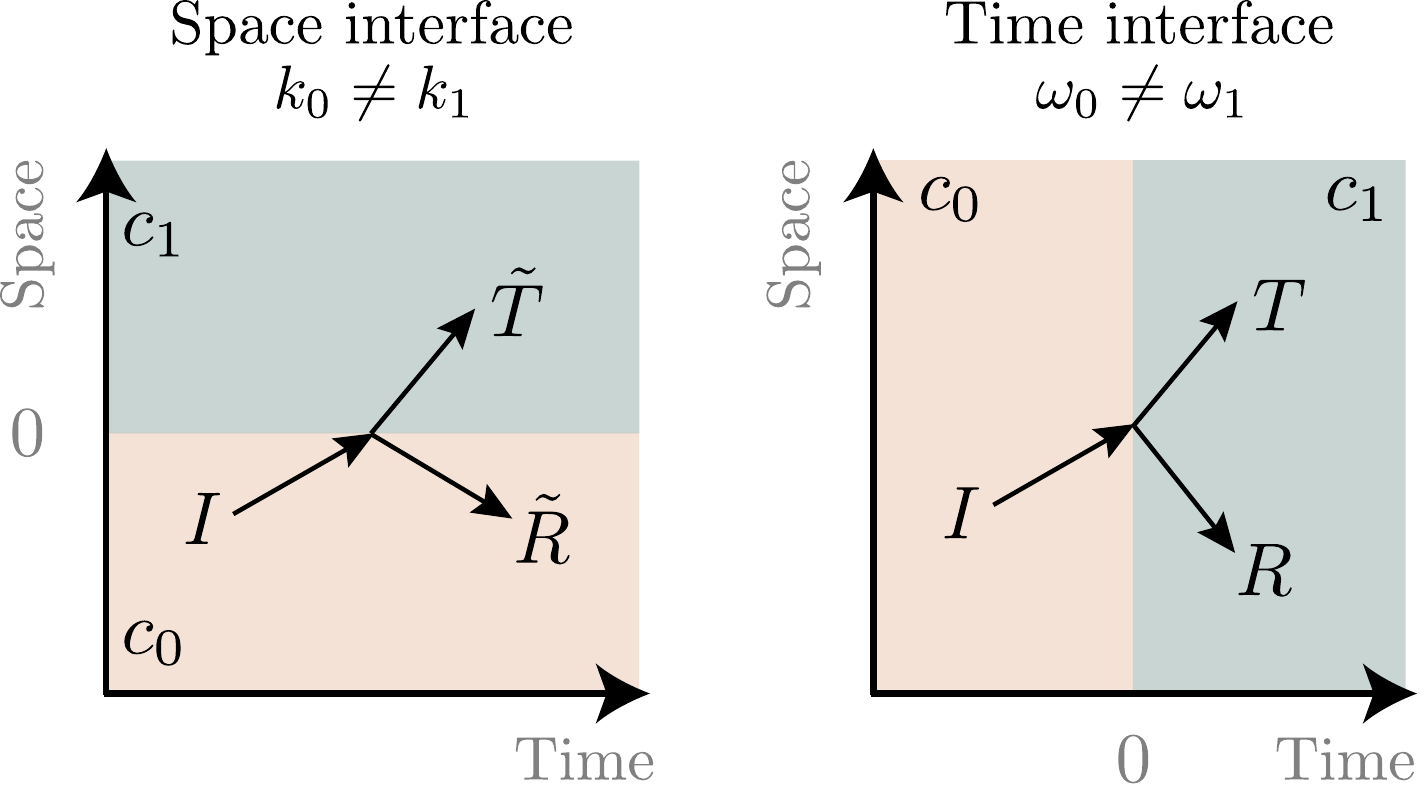}
\caption{Schematic of the wave scattering by a space interface versus a time interface.\label{time_interface}}
\end{figure}

Let us now consider a water wave governed by Eq. \eqref{anis_eq} and propagating in an unbounded medium which is modified everywhere in space at a given time instant $t=0$. When the medium switch is made abrupt enough, i.e. much faster than the wave period, the wave is scattered in time by splitting into two parts: a reflected backward-propagating wave and a transmitted forward-propagating wave \cite{shukla, tretyakov_temporal, kim_temporal_reflection_acoustic}. Since everything is homogeneous in space, the wavenumber is conserved at all times, in contrast to the classical problem of wave scattering by a space interface. Fig. \ref{time_interface} depicts two simplified schematics, reminding the difference between wave scattering by a space interface and a time interface (TI).  

It is straightforward to derive the continuity conditions at the TI starting from Eq. \eqref{anis_eq}, which are just the continuity of the field, $[\eta]_{t=0}=0$ and of its time derivative $[\partial_{t} \eta]_{t=0}=0$. The solution for $t<0$ is just the incident wave of angular frequency $\omega_0$, taking the form
\begin{equation}
\eta_{I}=\Re\{e^{i(k_{x} x+k_{y}y-\omega_{0}t)}\},
\end{equation}
while for $t>0$ the solution is composed of a reflected (backward) and a transmitted (forward) wave with new angular frequencies $\pm \omega_1$, which read as
\begin{equation}
\eta_{R}=\Re\{Re^{i(k_{x} x+k_{y}y+\omega_{1}t)}\}, ~ \eta_{T}=\Re\{Te^{i(k_{x} x+k_{y}y-\omega_{1}t)}\},
\end{equation}
with $R$ the reflection and $T$ the transmission coefficient. The continuity of $\eta(t=0)$ and $\partial_{t}\eta|_{t=0}$ yield in that order
\begin{equation}
1=R+T,
\end{equation}
\begin{equation}
\omega_{0}=\omega_{1}(T-R).
\end{equation}
Combining the two above relations we find the analytical expressions of the scattering coefficients: 
\begin{equation}\label{R_T_hachoir}
R=\frac{\omega_1-\omega_0}{2 \omega_1}, \quad T=\frac{\omega_1+\omega_0}{2 \omega_1}, 
\end{equation}
with $\omega_0$ and $\omega_1$ satisfying the dispersion relations of the two media at $t<0$ and $t>0$ respectively.  Considering the anisotropic medium of the previous section \ref{sA}, for an isotropic to anisotropic switch in the shallow water limit we would have $\omega_0=k \sqrt{g h^{+}}$ and $\omega_1=k \sqrt{g\left(h_x \cos ^2 \theta+h_y \sin ^2 \theta\right)}$, while in the reverse case $\omega_{0}$ and $\omega_{1}$ are swapped in Eq. \eqref{R_T_hachoir}. Furthermore, one can write down the angle for which $R=0$, which is known as the Brewster angle \cite{temporal_brewster}: 
\begin{equation}
\theta_{Br}=\sin^{-1}\left( \sqrt{(h^{+}-h_{x})/(h_{y}-h_{x})}\right). 
\end{equation}
\subsection{Temporal aiming}
\begin{figure}[h!tbp]
\centering
\includegraphics[width=1\columnwidth]{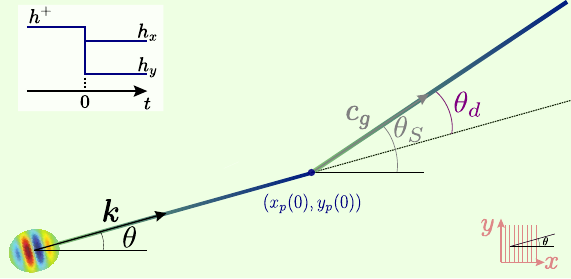}
\caption{Temporal aiming of a wavepacket using the metabathymetry: For $t<0$ the wavepacket travels at an angle $\theta$ in the isotropic medium where the fluid depth is constant everywhere in space. At $t=0$ when the wavepacket is located at position $(x_{p}(0),y_{p}(0))$ the plate array is lifted, giving rise to an effective anisotropic medium. Consequently, for $t>0$ the wavepacket travels at a new angle $\theta_{S}$, which is the angle of the energy flow and differs from $\theta$. The angle of deviation is denoted as $\theta_{d}$.  \label{temporal_waveguide}}
\end{figure}

In the case where at least one of the two media at a time interface is anisotropic we can achieve temporal aiming \cite{temporal_aiming}. That is because the anisotropy will cause the wavepacket to deflect from the $\boldsymbol{k}$ direction, while the time interface will trigger the deflection at a specific time. A schematic representation of the water wave temporal aiming is shown in Fig. \ref{temporal_waveguide}: A wavepacket is sent at an incident angle $\theta$ in an isotropic medium (constant water depth $h^{+}$) and at $t=0$ the medium switches to the effective anisotropic medium formed by the plate array, modifying the effective water depth along $x$ and $y$ ($h_{x}\neq h_{y}$). This medium change is achieved by vertically lifting the plate array at the fluid bottom, a movement that does not perturb the surface given that the plates are infinitely thin (see \cite{mei}). In this isotropic-to-anisotropic medium switch configuration, the wavepacket first travels at an angle $\theta$ for $t<0$ and after reaching position $(x_{p}(0),y_{p}(0))$ at $t=0$ then deflects from its initial trajectory, travelling at a new angle $\theta_{S}$ for $t>0$.  As already mentioned in section \ref{sA}, $\theta_{S}$ is the angle of the energy flow and is defined by the group velocity vector $\boldsymbol{c}_{g}$ (see Eq. \eqref{theta_S}). In the end, for $t>0$ the wavepacket is seen to deviate from the wavenumber path by an angle $\theta_d=\theta_S-\theta$, indicated in Fig. \ref{temporal_waveguide}. Importantly, a maximum angle of deviation $\theta_d$ can be found for a fixed anisotropy ratio $\mathcal{A}$, which occurs at the incident angle $\theta_{m}=\sec^{-1}\left(\sqrt{1+h_{x}/h_{y}}\right)$.
\section{Time-varying metabathymetry: Numerical analysis} \label{numerical_section}
In this section, in order to model the forthcoming experimental result, we numerically evaluate the propagation of the surface water wavepacket inside a medium which is rapidly transformed in time into a different one. We showcase a numerical example of the wavepacket scattering by a TI, by solving Eq. \eqref{anis_eq} using finite difference both in space and time, where the anisotropic tensor $\mathrm{H}=\mathrm{H}(t)$, swapping between isotropic and anisotropic (and vice versa) at $t=0$. We choose the carrier frequency $f_{0}$ which corresponds to $\omega l/\sqrt{g h^{+}}=0.68$ in the elliptic dispersion relation shown in Fig. \ref{effective_anis}(b) in the $(k_{x}l,k_{y}l)$ space, so as to guarantee a strong anisotropy ratio once the wavepacket is found in the anisotropic medium. We choose to work at the reference frame where the plate array is rotated at an angle with respect to the incident wave, thus the numerical results are represented at the coordinates $(X,Y)$, with $X=x\cos\theta+y\sin \theta$, $Y=-x\sin \theta+y\cos \theta$. The numerical analysis consists of solving a boundary value problem where the driving source at $X=0$ emits the signal
\begin{equation}
f_{c}(Y,t)=e^{-\frac{1}{2}\left(\frac{(t-t_{0})}{\sigma_{t}}\right)^{2}}\cos(2\pi f_{0} t)e^{-\frac{1}{2}\left(\frac{(Y-Y_{c})}{\sigma_{Y}}\right)^{2}},
\end{equation}
with $Y_{c}=25~\mathrm{cm}$, $\sigma_{Y}=7~\mathrm{cm}$, $t_{0}=0.4~\mathrm{s}$ and $\sigma_{t}=1/(2.5f_{0})$. When the length $\sigma_{Y}$ of the source is much larger than the carrier wavelength, a good directivity (narrow angular spread in $[X,Y]$ space) of the wavepacket is ensured.
\begin{figure}[h!]
\centering
\includegraphics[width=1 \columnwidth]{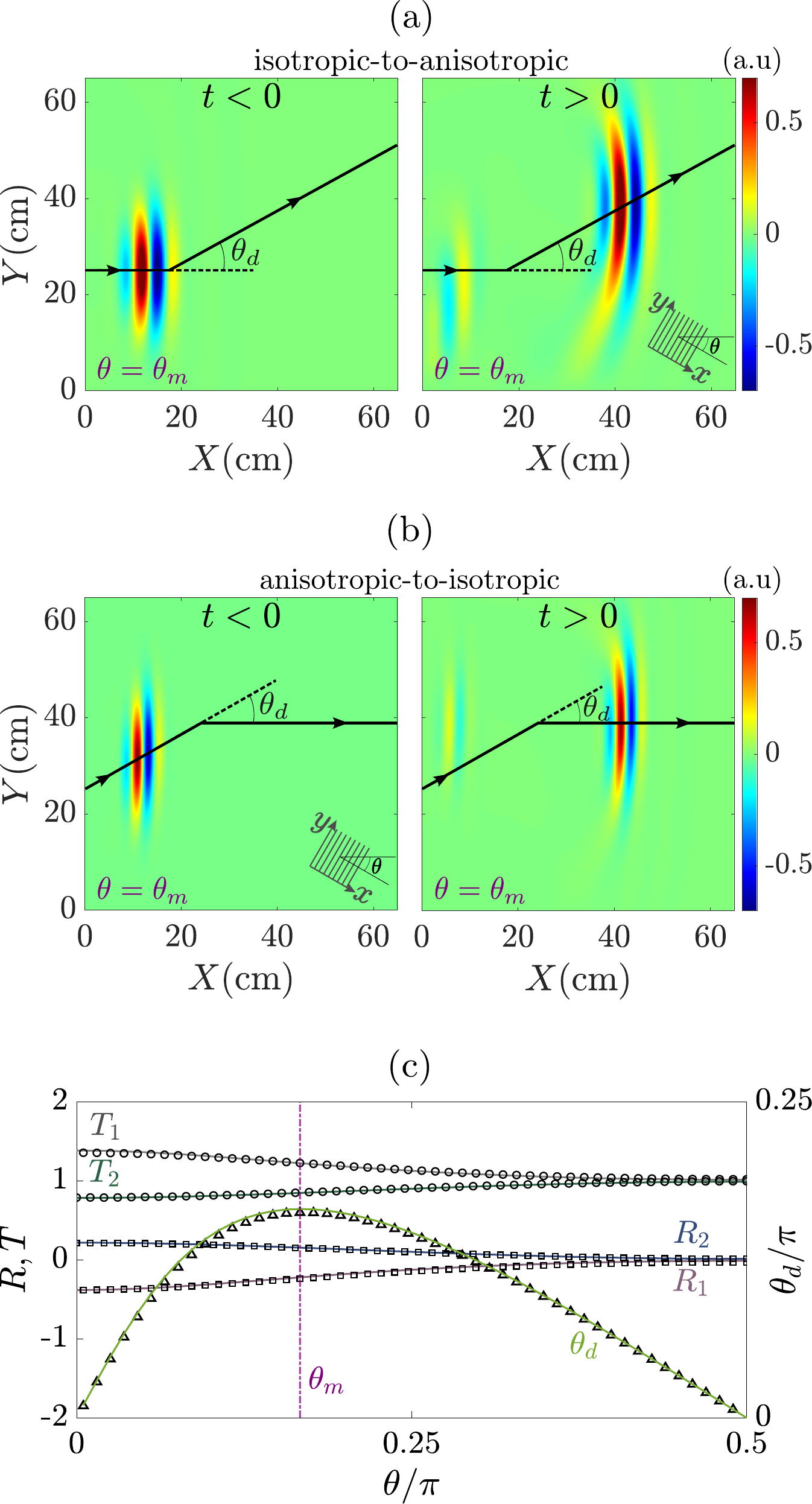}
\caption{Numerical example displaying the scattered fields from a TI for an isotropic-to-anisotropic switch in (a) and an anisotropic-to-isotropic switch in (b), when $\theta=\theta_{m}$. (c) $R$, $T$ and $\theta_{d}$ with respect to $\theta$ in the (a) and (b) scenarios of medium switch. Plain lines correspond to the theory (Eqs. \eqref{theta_S} and \eqref{R_T_hachoir}) and points to numerics. Here $l=0.8~\mathrm{cm}$, $\varphi=0.0625$, $h^{+}=2~\mathrm{cm}$, $h^{-}=0.5~\mathrm{cm}$, $f_{0}=6~\mathrm{Hz}$.\label{temporal_aiming_numerics}}
\end{figure}

Fig. \ref{temporal_aiming_numerics}(a) and (b) depict snapshots of the scattered wavepackets for an isotropic-to-anisotropic switch in (a) and anisotropic-to-isotropic in (b), where in both the incidence angle that produces the maximum deviation has been selected ($\theta=\theta_{m}$).  In the first example, portraying the surface profile before and after the TI (at $t=0$), one can recognise for $t>0$ the transmitted and reflected wavepackets of coefficients $T_{1}$ and $R_{1}$ accordingly and one can characterise the deflection of the wavepacket by tracking its center of mass $\mathrm{P}(t)=(X_{p}(t),Y_{p}(t))$.  At each time instant the center of mass coordinates is calculated in the numerical domain of surface $V$ from the expressions
\begin{equation} \label{center_of_mass}
X_{p}(t)=\frac{\int_{V} X\eta^{2}(X,Y,t)\mathrm{d}V}{\int_{V}\eta^{2}(X,Y,t)\mathrm{d}V}, ~ Y_{p}(t)=\frac{\int_{V} Y\eta^{2}(X,Y,t)\mathrm{d}V}{\int_{V}\eta^{2}(X,Y,t)\mathrm{d}V},
\end{equation}
yielding the black-lined trajectory of Fig.  \ref{temporal_aiming_numerics}(a). In the second simulation of Fig. \ref{temporal_aiming_numerics}(b), where the TI is also imposed at $t=0$, one can also perceive two wavepackets at $t>0$ which now have scattering coefficients $T_{2}$ and $R_{2}$. The trajectory followed by the forward propagating wave is depicted with the black line. Notice that even though the simulations in (a) and (b) start with the same incoming wavepacket of carrier frequency $f_{0}=\omega_{0}/2\pi$, the wavenumber is different in the two simulations owing to the different dispersion relations for $t<0$. Another conspicuous difference between cases (a) and (b) is that the reflected wave in (b) appears much weaker in amplitude. Finally, by iterating over all angles $\theta$ and computing for each one $T_{1}$, $T_{2}$, $R_{1}$, $R_{2}$ and $\theta_{d}$, we plot all the numerical results (in discrete black marks) along with the theoretically obtained curves (coloured lines) of Eq. \eqref{R_T_hachoir} in Fig. \ref{temporal_aiming_numerics}(c). The (2D+time) numerics are in very good agreement with the theoretical formulation. Furthermore, as previously commented, we indeed find an overall weaker reflection coefficient for the anisotropic-to-isotropic switch as compared to the reversed case, i.e. $|R_{2}|<|R_{1}|$. 
\section{Temporal aiming: Experiment in shallow water}
\subsection{Experimental setup}
\begin{figure}[h!tbp]
\centering
\includegraphics[width=1 \columnwidth]{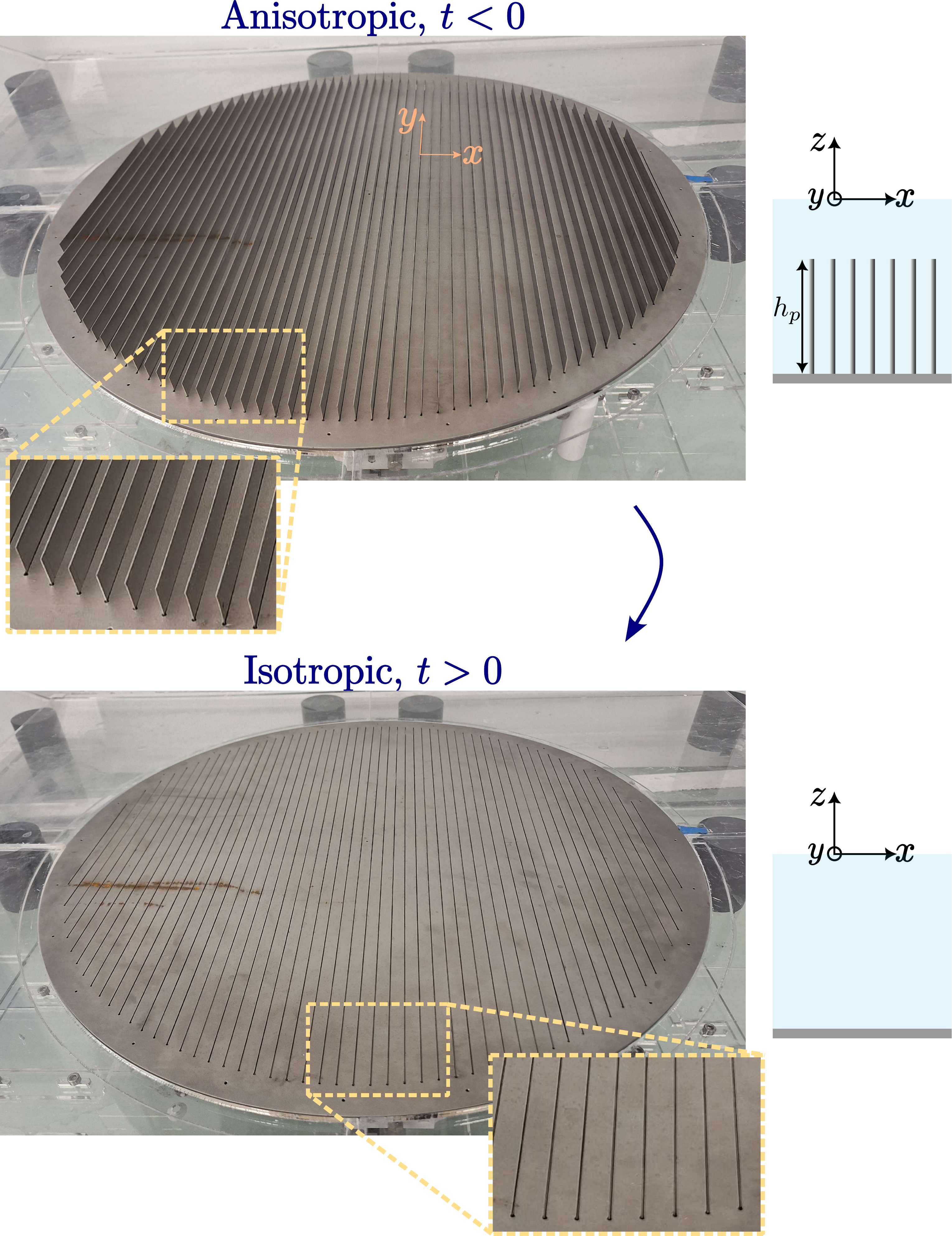}
\caption{Experimental setup showing the bottom profile before and after the switch at $t = 0$.\label{TV_experimental_setup}}
\end{figure}

For the experimental realisation of the temporal aiming we have designed the following setup: A circular disc of $55~\mathrm{cm}$ diameter and made of stainless steel is pierced into slits from where the plate array can pass vertically. This disc is placed at the center of a water tank of dimensions $120\times80~\mathrm{cm^{2}}$. Below the disc and inside the water tank there is a carefully designed mechanical system (see Appendix \ref{app_setup}) that allows the ascent and descent of the plate array through the disc, while being connected to an outside linear motor which initiates the movement. By imposing a prescribed jolt motion to the motor, the whole movement is transferred to the mechanical system which lifts (or drops) the plate array, while being isolated from the water surface. Fig. \ref{TV_experimental_setup} illustrates the top view of the experimental setup, specifically the topography change when we pass from the initial anisotropic medium to the isotropic one at $t = 0$. Finally, the plate array characteristics are $h_{p}=h^{+}-h^{-}=0.7~\mathrm{cm}$, with $h_{p}$ the plate height, $l=1~\mathrm{cm}$, $\varphi=0.05$, and the water depth is fixed as $h^{+}=1.45~\mathrm{cm}$.
\subsection{Wavepacket deviation}
 \begin{figure*}[t!]
\centering
\includegraphics[width=1 \textwidth]{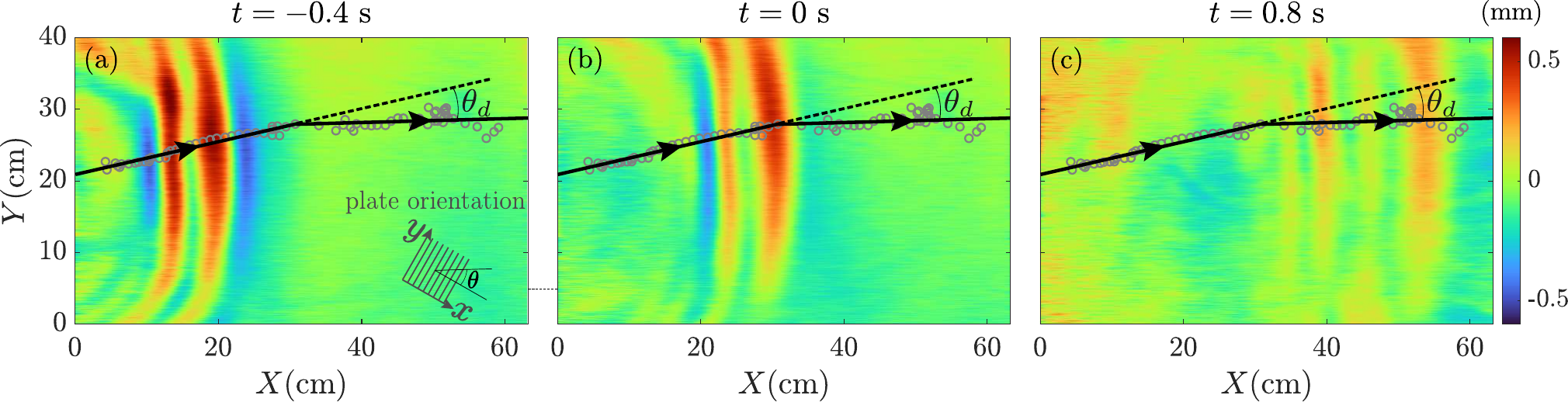}
\caption{Three different snapshots of the measured surface field before the change of medium in (a), at $t=0$ in (b) and after the change of medium in (c). The measurements portray the wavepacket being deflected by an angle $\theta_{d,exp}=11.77^{\circ}$. \label{TV_snapshots}}
\end{figure*}
In our experiments, we concentrate on examining the wavepacket deviation when we switch the initial anisotropic medium to the isotropic one at $t=0$. This choice is motivated on one hand by the fact that reflection is minimal in this configuration, allowing the focus to be solely on the transmitted wave, and on the other hand from the easier and more controlled manipulation of our experimental setup in the downward plate motion. The experiments are conducted for a plate array that is inclined with respect to the wave vector, such that $\theta=31.27^{\circ}$ (refer to Fig. \ref{temporal_waveguide} for the illustration in the plate array coordinates). A cylindrical wavemaker of $20~\mathrm{cm}$ length acting on the water surface is used in order to generate a wavepacket with bandwidth $4.55\pm1.25~\mathrm{Hz}$. Once the wave arrives at a given position in the anisotropic medium the plate array is dropped within $0.2~\mathrm{s}$, which is the imposed jolt motion duration of the motor. 

In each experiment the surface elevation $\eta(X,Y,t)$ is measured using the Fourier transform profilometry technique \cite{pablo_FTP, ftp_surface}, which is a space time resolved optical method based on recording the deformation of a sinusoidal pattern projected at the water surface and translating it to the water surface deformation. Then, the angle of deviation $\theta_{d}$ of the wavepacket is evaluated for each experiment, by following its center of mass before and after the medium switch, as was done in the numerical simulations. Fig. \ref{TV_snapshots} depicts an experimental measurement where we can track the wavepacket in time, with the black trajectory being calculated from fitting the $(X_{p}(t),Y_{p}(t))$ wavepacket coordinates at the consecutive snapshots in time (illustrated with the gray circles), with time step $\Delta t=0.033~\mathrm{s}$ and time interval $t\in[-1.27, 1.13]~\mathrm{s}$. Note that if one considers the shortest wavelength in the wavepacket spectrum $\lambda_{\mathrm{min}}$ (for frequency 5.8Hz), which is approximately $4.5~\mathrm{cm}$, it can be verified that the gravity regime (neglecting capillary effect) assumed is reasonable since the capillary wavelength in room temperature conditions is $\lambda_c=1.7~\mathrm{cm}\ll \lambda_{\mathrm{min}}$. For this wavelength $\lambda_{\mathrm{min}}$, it is also found that the shallowness parameter in the isotropic medium is $\omega\sqrt{h^{+}/g}\sim1.4$, while the microstructure shallowness parameter is $\omega \sqrt{l/g}\sim1.16$. The fact that these ratios are of the order one suggests that weak dispersive effects are to be expected for the higher frequencies of the spectrum both for the wave propagation in the isotropic and the anisotropic medium. Even though this behaviour is observed in Fig. \ref{TV_snapshots}, the wavepacket can still be efficiently tracked in time and information on the angle of deviation can be extracted. The post treatment of the experimental data gives us an average value for the deviation angle which is $\theta_{d,exp}=(11.77\pm0.46)^{\circ}$. This value is calculated from a total of four experiments (see Appendix \ref{sup_exp}) and is consistent with the theoretical one, $\theta_{d}=11.42^{\circ}$, with a relative error of $3\%$. 

An additional remark to be made is that we do not observe a reflected wave, something which is consistent with the weak reflection predicted by the theory for the anisotropic-to-isotropic switch (discussed in section \ref{numerical_section}). Also to be noticed is the wave attenuation that is due to two main sources of losses in the experiment: firstly, viscous friction originating from the bottom boundary layer since we are in shallow water and secondly from the surface contamination over time which forms a viscoelastic surface film altering the surface tension and increasing the damping rate compared to a clean-surface case (see \cite{marangoni,miles}). Finally, while the experimental execution specifically tackles the anisotropic-to-isotropic change, the isotropic-to-anisotropic change can also be conceivable as long as sufficient force is injected with a motor to lift the plate array and also provided that there is no friction of the upward moving plate array with the pierced disc.
\section{Conclusion}
In this study we presented the experimental realisation of the water wave analog of the temporal aiming in the long wavelength approximation using a time-varying metabathymetry.  Since a metamaterial plate array is an efficient tool to create a 2D effective anisotropic medium in shallow water, we utilise it along with time variation to pass from an isotropic initial medium to an anisotropic one and vice versa. The key advantage with an array of infinitely thin plates is the fact that its vertical movement does not perturb the surface, making it a suitable candidate for applications involving rapid medium changes in time, such as time interfaces. In this spirit, by sending a surface wavepacket over an initially constant bathymetry and then suddenly lifting the plate array at the fluid bottom so as to create an anisotropic metabathymetry, we can initiate wavepacket scattering by a time interface and subsequently modify the propagation direction thanks to the anisotropy. The time interface can be ensured when the duration of the medium switch is much smaller than the wave period, and during this process, the wavenumber is conserved. 

By applying the continuity conditions at the interface we derived explicit expressions for the scattering coefficients and the angle of deviation. This simple theoretical modelling shows very good agreement with the numerical simulations performed for the full 2D anisotropic wave equation for all incident angles, with the reflection coefficient being weaker for an anisotropic-to-isotropic change compared to the reverse process. Exploiting this last fact and with the main focus being the transmitted wave, we designed an experimental setup optimised for an anisotropic-to-isotropic change, allowing the plate array to descend abruptly at a given time. From the experimentally measured surface fields we pinpointed the wavepacket trajectory in time and we extracted its deviation angle, which is quite consistent with the theory. 

We believe that this work, which highlights the experimental adaptation and realisation of time interface control of a water wavepacket with temporal aiming, could be followed by other experimental characterisation in acoustic, mechanical or photonic setups. Particularly in acoustics, thin plate structures which are periodically arranged can also engineer an effective anisotropic medium. Such a medium combined with a fast lifting mechanism for the plates could potentially lead to an experimental temporal aiming concept. Overall, challenges could mainly arise in terms of triggering fast enough medium changes (compared to the typical time scale of the signal) and also on finding suitable mechanisms that drive the medium switch without them acting as additional sources.
\begin{acknowledgments}
The authors acknowledge the support of the ANR under grant no. ANR-21-CE30-0046 CoProMM. We further acknowledge the contributions of the engineering team of the PMMH laboratory, Laurent Quartier, Xavier Benoit Gonin and Amaury Fourgeaud for their input and technical support in the construction of the experimental setup. 
\end{acknowledgments}

\appendix*

\section{Mechanical design of the time-varying experiment}\label{app_setup}
\begin{figure}
\centering
\includegraphics[width=.55 \textwidth]{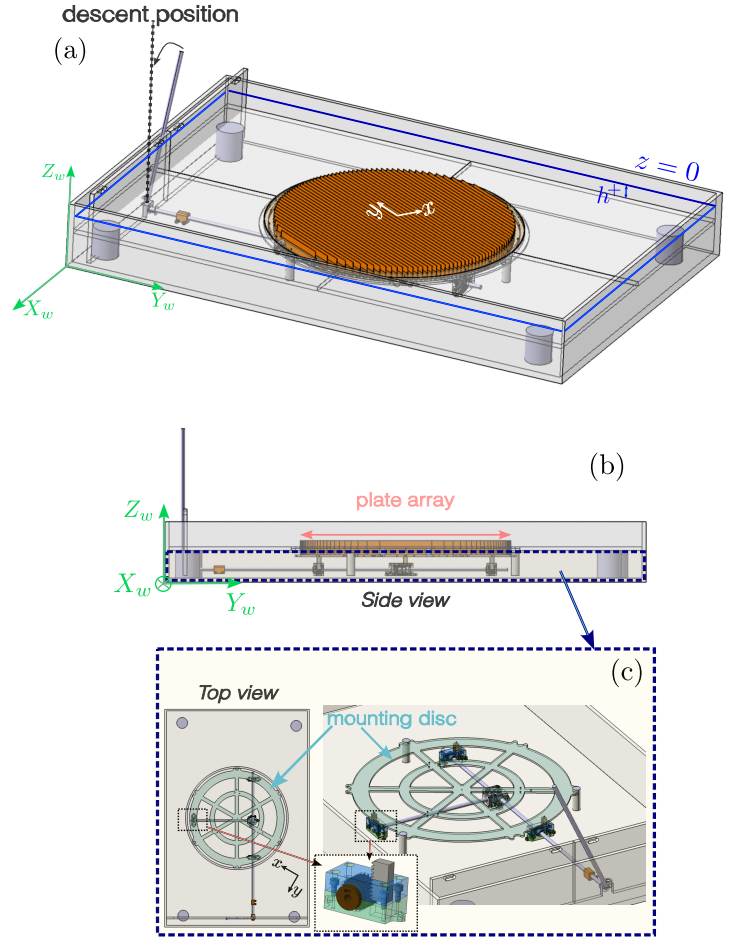}
\caption{(a) Full experimental setup, consisting of the water tank, the plate array and the mechanical compartments responsible for the vertical movement of the array. The rod, which initiates the downward plate movement, rotates in order to activate the descent of the plate array by means of a mechanical system positioned below the level $z=-h^{+}$.  (b) Side view in the $(Y_{w},Z_{w})$ plane of the experimental setup shown in panel (a). (c) Supplementary schematic representation of the mechanical system below the level $z=-h^{+}$ without the plate array: The mounting disc which supports the plate array descends when the rod turns to its descent position. Experimental setup images courtesy by Laurent Quartier. \label{fig_appendix}}
\end{figure}

In this section we provide a more detailed description of the experimental setup, focusing on the mechanical aspects driving the downward movement of the plate array. As was previously described, the plate array passes from a pierced stainless steel disc that occupies a circular space of $55~\mathrm{cm}$ diameter at the center of a water tank of dimensions $120\times80~\mathrm{cm^{2}}$. The plate array has been assembled using stainless steel plates of $0.5~\mathrm{mm}$ width and small magnets of $9.5~\mathrm{mm}$ height attached between each plate and its neighbour, so that the array periodicity is $l=1~\mathrm{cm}$. The movement of the plate array is independent of the pierced disc, which always remains fixed at its position, and is driven by a mechanical system beneath the disc which in turn is connected to a rod extending outside the water tank (see Fig. \ref{fig_appendix}(a)). The rod can be pushed in the $X_{w}$ direction from an initial position to a new descent position in order to drop the plate array at the fluid bottom (at $z=-h^{+}$) through this mechanical system.
 \begin{figure*}[t!]
\centering
\includegraphics[width=1 \textwidth]{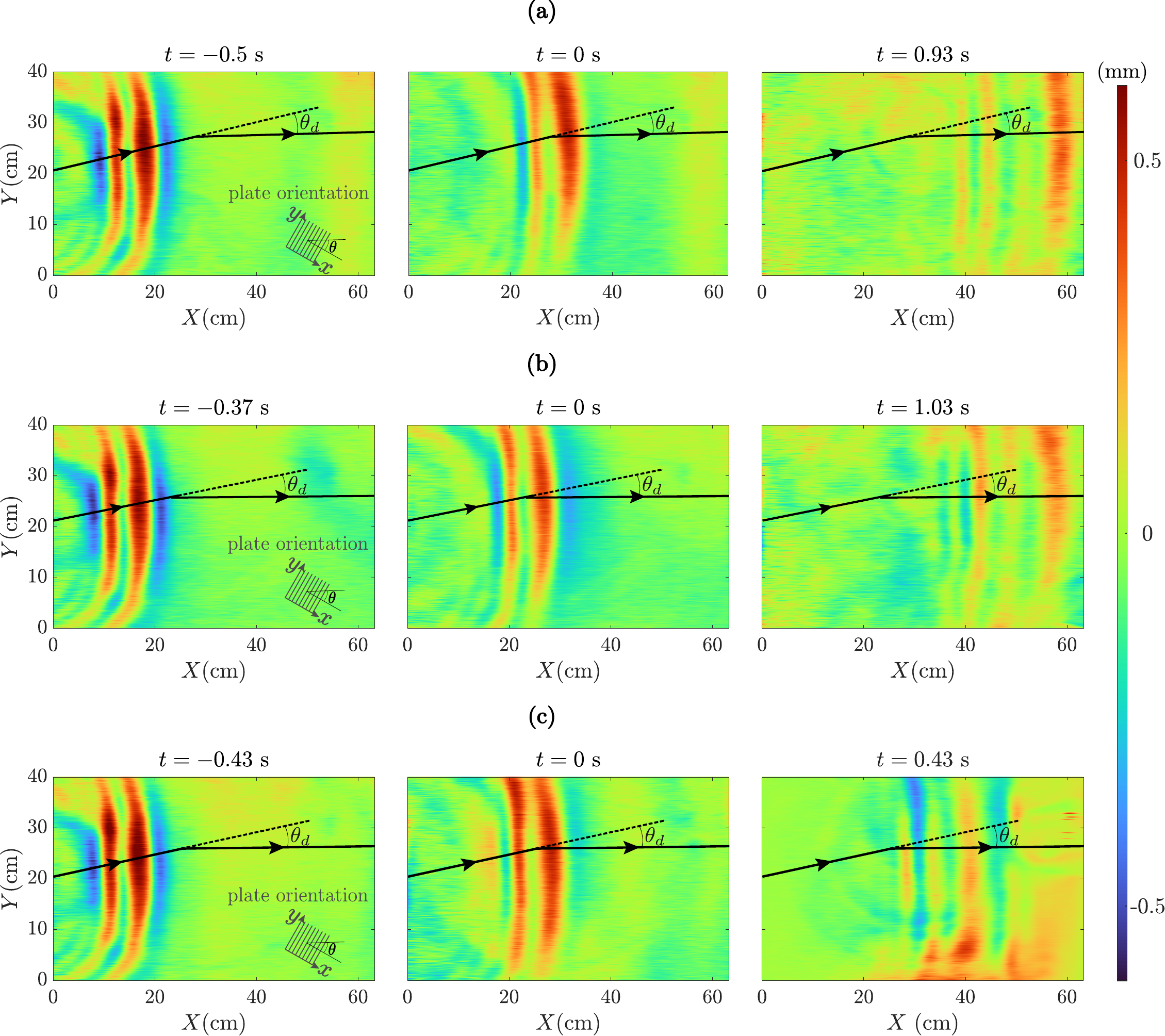}
\caption{(a-c) Experimental demonstrations of the wavepacket deviation, where $t=0$ is defined in each experiment as the time where the plate array is dropped. \label{TV_snapshots1}}
\end{figure*}
In Fig. \ref{fig_appendix}(a) the system is shown in the upward position, depicting also the position of the rod, while in Fig.  \ref{fig_appendix}(b) panel (a) is represented from the side view, i.e. in the $(Y_{w},Z_{w})$ plane defined in panel (a), where the mechanical system below the plate array is more visible. The latter system is composed of another thin circular mounting disc which supports the plate array and a guiding structure made of three cylindrical pillars fixed to the bottom of the water tank from where the downward movement of the mounting disc is guided. The components of the mechanical system along with the rod are illustrated on their own in Fig. \ref{fig_appendix}(c). The principle of the movement is that when the rod rotates it triggers the downward motion of the mounting disc via some connecting compartments, one of which is also shown in close-up view on panel (c). Finally, the entire movement of the mounting disc with the plate array is performed below the reference fluid bottom of our experiments (at $z=-h^{+}$) so that it does not perturb the free surface.
\section{Supplementary experiments}\label{sup_exp}
The wavepacket deflection from its initial path can be appreciated from three additional experimental demonstrations (a)-(c) in Fig. \ref{TV_snapshots1}. Each panel, corresponding to a distinct experiment, depicts snapshots of the surface field at different times, in the same spirit as in Fig. \ref{TV_snapshots}. The time $t=0$ is set for each experiment separately as the time where the medium is switched from anisotropic to isotropic, the two media being identical in all experiments. In that sense, the medium switch can occur earlier or later in the wave propagation during the time window $t<0$. For instance, one can notice that the wavepacket on panel (a) at time $t=0$ has reached a different position in space than the wavepacket on panel (b) at $t=0$. This leads to slightly different trajectories in space traced by the wavepacket, something which further illustrates the concept of temporal waveguiding.

%\clearpage
\bibliographystyle{apsrev4-2}
\bibliography{temporal_aiming_bibliography}% Produces the bibliography via BibTeX.

\end{document}